\documentclass[12pt]{iopart}

\usepackage{iopams}  
\usepackage{graphicx}
\usepackage{srctex}
\usepackage[dvipsnames]{xcolor}

\begin{document}

\title[]{Numerical studies for an {\bfseries\emph{ab initio}} investigation into the Boltzmann prescription in statistical mechanics of large systems}

\author{V.\ Dossetti}
\ead{victor.dossetti@correo.buap.mx, vdossetti@gmail.com}
\address{CIDS-Instituto de Ciencias, Benem\'erita Universidad Aut\'onoma de 
Puebla,\\ Av.\ San Claudio esq.\ 14 Sur, Edif.\ IC6, Puebla, Pue.\ 72570, Mexico}

\author{G.M.\ Viswanathan}
\ead{gandhi.viswanathan@gmail.com}
\address{Departmento de F\'{\i}sica and National Institute of Science and Technology of Complex Systems, Universidade Federal do Rio Grande do Norte, Natal--RN, 59078-970, Brazil}

\author{V.M.\ Kenkre}
\ead{kenkre@unm.edu, vmkenkre@gmail.com}
\address{Department of Physics and Astronomy, University of New Mexico,\\ Albuquerque, NM 87131, USA}

\begin{abstract}
We present numerical investigations into the question of the validity of the Boltzmann prescription in Statistical Mechanics for large systems, addressing the issue of whether extensivity of energy implies the extensivity of the Boltzmann entropy. The importance of the question stems from the fact that it is currently considered open by some investigators but quite settled by others. We report {\it ab initio} results for gas-like Hamiltonian systems with long-range as well as short-range interactions, based on simulations that explicitly consider more than $2^{30} \approx 10^9$ states of the full Hilbert space. The basis of the technique is Monte Carlo algorithms. Despite the largeness of the numbers used, careful inspection shows that the systems studied are still too small to settle uniquely the issues raised. Therefore, the new approach outlined represents a first step in addressing on first principles the question of non-extensive statistical mechanics. General theoretical comments are also supplied to supplement the numerical investigations.
\end{abstract}

\vspace{2pc}

\maketitle

\section{Statement of the problem}

The statistical mechanics of non-extensive systems is still considered an open problem \cite{bal20, rib21}. On the one hand, many studies in the past three decades have addressed the issue of generalizations of Boltzmann-Gibbs statistical mechanics \cite{tsallis1, tsallis-review, tsallis-review19, abe1, abe2, abe3, rus16a, rus16b}. These studies typically assume a non-logarithmic entropy, leading to power law distributions (e.g., Zipf-Mandelbrot distributions \cite{abe1,abe2,abe3}) instead of exponential distributions of energy. Tsallis statistical mechanics falls into this category and it has found many applications throughout the years \cite{cur97, ish15, cig18, gup18, wen19, kla20, dep20, kol20, dep21, oli21}.  On the other hand, not everybody is fully convinced about the validity of such approaches to statistical mechanics \cite{science, lim20}.

It is well known that for gas-like systems governed by Hamiltonians with only short-range interactions between the constituent subsystems, the total energy is extensive, i.e., additive, in the limit of infinite subsystems (and reservoirs). Moreover, the Boltzmann entropy is also necessarily extensive because subsystems far enough apart from each other act independently. Indeed, the Boltzmann factor is the unavoidable consequence of the extensivity of the Boltzmann entropy. But what happens if the Hamiltonian contains long-range interactions that decay with distance as a power law? This question lies at the heart of the entire problem.

One way of formulating the problem succinctly is as follows. Given a gas-like Hamiltonian system of macroscopic size, with interactions that are not only of short range but can also be of long range between the many separate components, what is the functional form of the distribution of energy $E_S$ for a macroscopic subsystem? Is it proportional to $\exp[-\beta E_S]$ as we have learned traditionally from Boltzmann arguments or is it different, perhaps  an exotic power law tailed distribution as suggested by Tsallis and others? 

There have been two different attempts to justify non-Boltzmann distributions. One approach appeals to myriads of experiments  that show power-law scaling and thus consists of  an empirical justification \cite{tsallis-review19}. Some have argued against  such arguments along with the claim that power law scaling in systems far from thermodynamic equilibrium has little to say about systems in equilibrium. While these objectors do not deny power-law scaling in the Gutenberg-Richter law for earthquakes \cite{gut56}, for example, they argue that  tectonic plates are not in thermodynamic equilibrium, and that such arguments do not contribute to settling issues about equilibrium. They assert that direct and convincing empirical evidence supporting Tsallis and similar distributions for systems in equilibrium is still lacking. In reacting to similar arguments that treat Tsallis statistics as the correct description not for systems in equilibrium, but rather for metastable states \cite{cam02, plu07, liv10, cir15}, the objectors point out that  the approach to equilibrium is a separate topic and metastable states provide no compelling basis to abandon the Boltzmann entropy.  

Abe and Rajagopal have attempted to construct a theoretical foundation for Tsallis statistics \cite{abe1,abe2,abe3}.  In brief, their argument proceeds by (i) assuming a Zipf-Mandelbrot distribution with power law behavior for the energies, (ii) showing that Khinchin's derivation in terms of Gaussian distributions and the Central Limit Theorem must be generalized via L\'evy $\alpha$-stable distributions and the L\'evy-Gnedenko Generalized Central Limit theorem, and finally (iii) recovering the usual thermodynamic properties relating energy, temperature, and entropy.

A similar argument is to allow different parts in the total system to maintain long-range correlations in their energies, so that the convergence to Gaussian distributions demanded by the Central Limit Theorem fails. The expected convergence would be disrupted by the long-range correlations in energy, which violate the conditions for the theorem to hold.  The Tsallis factor might be justified if the energy correlations lead to a $q$-Gaussian distribution of energy for the total system. Both these approaches assume, implicitly if not explicitly, that the interaction energy is non-negligible even for arbitrarily large separations between constituent components.  In other words, both these arguments assume strongly interacting subsystems.

Because there is no denying that, if justified, alternatives to the Boltzmann prescription would be exciting to pursue, we undertook a direct numerical investigation into some systems that we describe below. First we touch upon analytical underpinnings of the Boltzmann prescription as we have learned traditionally in textbooks. Then we initiate a computational study of large systems with long-range interactions that might, at least in principle, possibly produce departures from the Boltzmann extensivity result. Finally, we conclude the presentation with a few remarks.

\section{Analytical preliminaries}

Let us begin by reviewing what is well known in the traditional development of statistical mechanics as laid out in innumerable texts \cite{rei16, hua91, pat22, jac00, wid02, fey98, kub71}.

If you a toss a coin 20 times up in the air, it might come down 6 times heads and 14 times tails.\footnote{The experiment is, needless to state, equivalent to tossing 20 non-interacting coins and counting the outcome as well.}  If, however you toss the coin $N$ times where, rather than $N=20$, you have $N \gg 1$, you will find that the probability of heads and tails is about equal. In other words, equal \emph{a priori} probability law will hold for large $N$. The law dictates, thus, that the probability is $1/\Omega$ where $\Omega=2$ is the number of states available. Precisely the same law is assumed for an isolated-system situation (\emph{microcanonical} ensemble case) in statistical mechanics where, if energy of the system is denoted by $E$, then the probability that the system has that energy is given by $p(E)=1/\Omega(E)$. Numerous attempts to justify the equal \emph{a priori} probability law have been made by investigators in ergodic theory contexts. However, it is difficult to use them confidently because of the problem posed by the requirement of metrical transitivity that has been very rarely shown to be satisfied. It is best to avoid the ergodic arguments and to take, as many have, (see refs.\ \cite{rei16, hua91, pat22, jac00, wid02, fey98, kub71}) the principle of equal \emph{a priori} probability as a law of nature applying to systems in thermal equilibrium.\footnote{Note here, for instance the words of Feynman on page 1 of ref.\ \cite{fey98} and of Kubo on page 7 of ref.\ \cite{kub71}.}

Let a supersystem (merely meaning a very large system)  be isolated. Let us focus on the probability $p(E_S)$ that a part of it, we will call it the system, has energy $E_S$. Let us name the rest of the supersystem the reservoir that works as a thermal bath for the system. Generally, if the energy of the bath is $E_B$, and the interaction energy between system and bath is $V$, the energy of the super system is $E_S+E_B+V$.

Let us use, as an approximation, the neglect of $V$. This is assumed to be the result of the nature of the system (motivated and justified by the number of degrees of freedom being enormous and the range of interaction not being too long). The approximation allows us to conclude that 
\begin{equation}
p_1(E_S)p_2(E_B)=p_{1,2}(E_S+E_B).
\label{productandsum}
\end{equation} 
Here we are considering two systems (1 and 2) in turn as separate subsystems on the left side of the equation, and the two together as a single subsystem (1,2) on the right of the equation. Without any further arguments, one can then conclude that $p$ is an exponential function of energy: the conclusion is assured by the energies being sums and the probabilities being products, simultaneously. Both are consequences of the neglect of $V$. There are no possibilities of Tsallis-like departures from the Boltzmann prescription if $V$ is neglected and the product condition of probabilities is assumed.

The argument is straightforward and can be appreciated from discussions in all the numerous texts quoted.\footnote{It is wise to take care not to be misled by occasional discussions encountered in the literature that suggest Taylor expansions of the exponential of a logarithm along with the statement that logarithms are dictated by the large number of degrees of freedom.} Standard graduate texts such as by Reichl \cite{rei16}, Huang \cite{hua91}, and Pathria \cite{pat22} have provided this explanation for generations of readers learning the principles of statistical mechanics for the first time. We recommend to the reader particularly the lucid discussions in the three additional texts \cite{jac00, wid02, fey98} where we believe the development is crystal-clear and devoid of possible confusions. 

Although the neglect of $V$ in systems of enormous numbers of degrees of freedom assures one of the Boltzmann prescription (i.e., the exponential probability law), we believe that the need remains to investigate its validity for systems in which $V$ cannot be neglected, for instance in systems with long-range interactions \emph{even if} the system size is huge. The essence of the present study is that we attempt to do in this in a direct computational manner, taking advantage of another well-discussed idea that has, however, not been used for this specific purpose, to the best of our knowledge. The idea is based on  relating the probability $p(E_S)$ of the system to have energy $E_S$ to properties of the bath having the energy $E_B = E - E_S - V$. In this particular expression, $E$ is the (constant) energy of the supersystem that is the combination of the system and the bath and that is considered to be itself isolated and therefore with a constant $E$.\footnote{Note that this has little connection with Eq.\ (\ref{productandsum}) whose assertion addresses two parts of the supersystem considered first separately and then together.} 

The probability $p(E_S)$ of the system to have energy $E_S$ is given by the probability of the bath (reservoir) to have the energy $E_B = E - E_S - V$. Whereas the standard textbooks \cite{rei16, hua91, pat22, jac00, wid02, fey98, kub71} assume the neglect of $V$ in this expression for $E_B$, we set ourselves the task of \emph{calculating} this quantity explicitly when $V$ cannot be necessarily neglected. We do this numerically from independent considerations of the reservoir Hamiltonian in the presence of arbitrary-range interactions $V$. If $V$ effects are negligible, we return to Boltzmann. If $V$ is not negligible, we get an opportunity to look for departures as suggested by Tsallis and his collaborators. 

We have found that our procedure that we report below, shares the philosophy set out in the two simple examples given by Feynman in his text \cite{fey98} where (on pages 3-5 of the book), he has taken the reservoir to consist, in the first case, of N independent harmonic oscillators and, in the second, of N noninteracting particles in a box. Interactions $V$ are neglected in that analysis and the canonical distribution is explicitly verified. Although similar, our study has as its aim a search for departures from the canonical distribution that might arise from long-range interactions. Therefore, by contrast, what we study below is spin systems first of short range and then of intermediate and long range.

\section{Numerical results}

The number of states $\Omega$ grows  with the system size. Hence, most if not all numerical studies have used sampling methods, e.g.\ Monte Carlo simulations using Glauber or Metropolis algorithms to ensure detailed balance \cite{tur11, wal15}. Many such methods assume the Boltzmann factor. 

To overcome such limitations, we run {\it ab initio} simulations without assuming the Boltzmann factor. For this, we sweep the entire Hilbert space of the total system (supersystem) using brute force to obtain the distribution $\Omega(E)$. Then, we select just those states that correspond to the microcanonical ensemble (for a given energy $E$ or, in our case, a small window of energies around $E$ as explained below), and use this ensemble for the supersystem to study the corresponding ensemble for the system and its distribution $p(E_S)$. As a test subject for our numerical studies, we consider the one-dimensional Ising model with periodic boundary conditions.

\subsection{Numerical estimation of $\Omega(E)$}

\begin{figure}[t]
\includegraphics[width=0.44\textwidth] {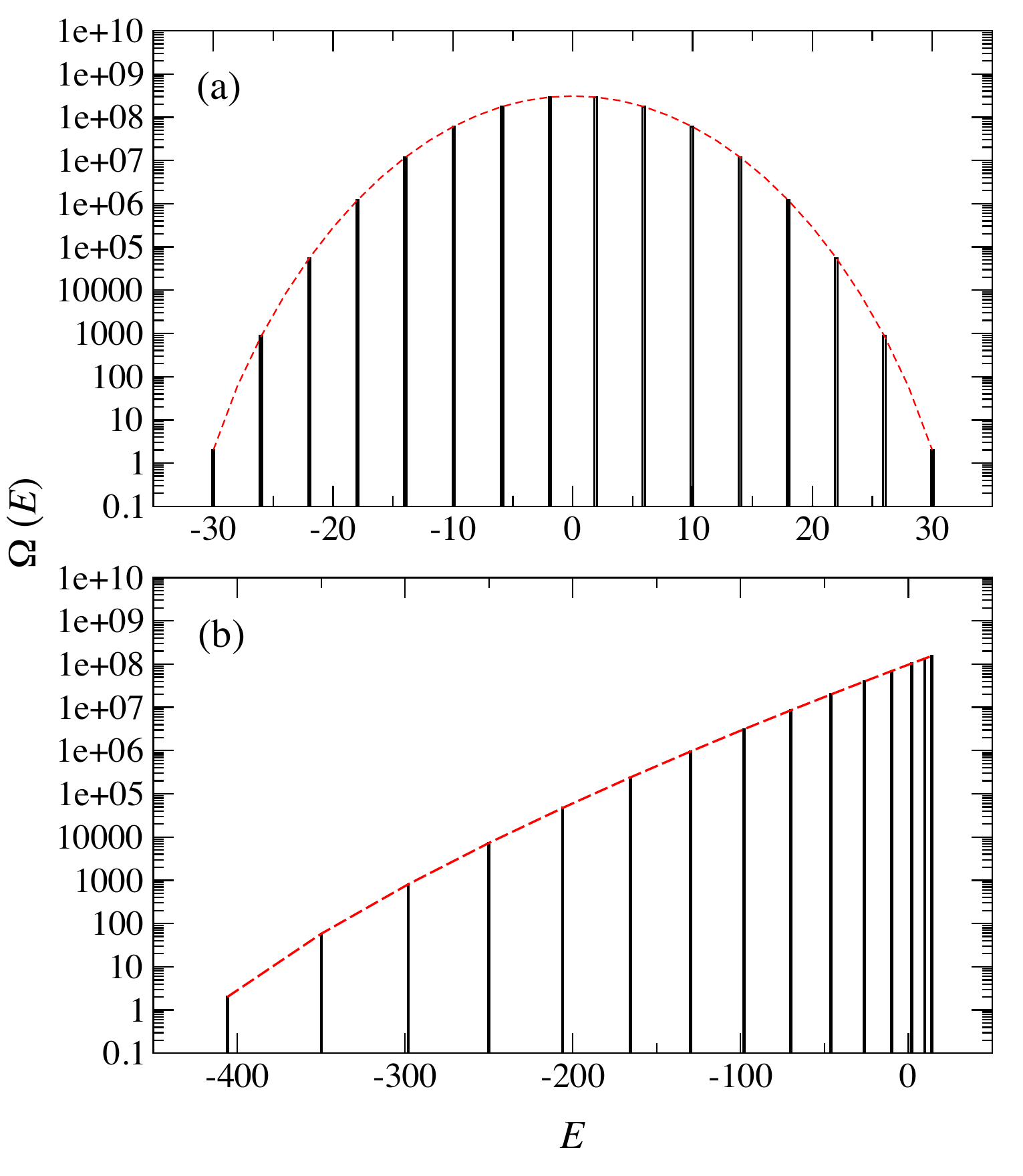}
\caption{Comparison of analytically and numerically estimated distribution of energy $\Omega(E)$ for two illustrative examples: (a) nearest neighbor 1-D Ising model with $N=30$ spins and (b) mean field Ising model with $N=29$ spins. The even and odd numbers of spins were chosen due to symmetry considerations. We see perfect agreement between the analytical and numerical results.}
\label{fig-omega}
\end{figure}

Prior to investigating the statistical mechanical properties of extensive and non-extensive systems, we first need to know how the number of states of the supersystem depends on the energy. Figure \ref{fig-omega} compares the numerically estimated values of $\Omega(E)$ for the nearest-neighbor 1-D Ising model and the mean field Ising model, which are both easy to test.  For the one-dimensional Ising model with nearest neighbor ferromagnetic interactions, \mbox{$H=-J\sum^N_i s_i s_{i\pm 1}$,} with periodic boundary conditions, where $s=\pm 1$ are the ``spins,'' and the constant $J>0$ has units of energy.  We have taken $J=1$ without loss of generality.  For the mean field model, all spins interact with all other spins: $H=-J\sum_{ij} s_i s_j$.  

The histograms for $\Omega(E)$ (black curves in Fig.\ \ref{fig-omega}), for the nearest neighbor and mean field cases, were numerically obtained by generating all of the configurations of the system and calculating the energy of each state though the hamiltonians provided in the previous paragraph. The analytical curves in Figure \ref{fig-omega} were estimated as follows.  

For the 1-D nearest neighbor Ising model, the ground state is two-fold degenerate, due to the spin inversion symmetry. There are $2N$ states corresponding to the first excited level, which arises when there are two domains separated by a single domain wall. The $n$-th excited level $E_n$ corresponds to
\begin{equation}
\Omega(E_n) = 2  {N \choose n} = 2 {\Gamma(N+1)
\over \Gamma(N-n+1) \Gamma(n+1)}   \;\; ,
\label{eq-ising}
\end{equation}
number of states. The energy depends on $n$ as
\begin{equation}
E(n) = 2nJ-N \;\; ,
\end{equation}
so
$E_n$ and $n$ are linearly related.

For the mean field model, one again recovers Equation (\ref{eq-ising}), by interpreting $n$ not as the number of domains but rather as the number of flipped spins.  The energy, however, is no longer linear in $n$, instead the energy is given by 
\begin{equation}
E(n) = - {(N-n)(N-n-1)\over2} - {n(n-1)\over2} + {n(N-n)} \;\;, 
\end{equation} 
where the first and second terms represent the interactions between the ``up'' and ``down'' spins respectively, while the last term represents the interaction between the set of ``up'' spins with the set of ``down'' spins.

We see perfect agreement between the analytical and the numerical results.

\subsection{From microcanonical to canonical} 

In order to consider systems with interactions of any range, the \hbox{models} provided above can easily be generalized as
\begin{equation}
H=-J\sum_{ij} s_i s_j w_{ij}(|x_i - x_j|),
\label{eq-gen-hamiltonian}
\end{equation}
where we have introduced the weight function $w_{ij}$ that depends on the distance between to given spins. For simplicity, the distance between two consecutive spins is taken to be $|x_i - x_{i\pm 1}| = 1$. In this way, for example, \hbox{$w_{ij}(|x_i-x_j|) = \delta_{i \, i\pm 1}$} for the nearest neighbor model, where $\delta_{ij}$ is the Kronecker delta, while  \hbox{$w_{ij}(|x_i-x_j|) = 1$} for the mean field model. For interactions with a varying range, we introduce a power law weight function that decays with the distance, i.e.,
\begin{equation}
w_{ij}(|x_i - x_j|) = \frac{1}{|x_i - x_j|^\gamma},
\label{eq-pow-law}
\end{equation}
where the parameter $\gamma$ is used to control the range of the interaction. As such, we are now dealing with a \hbox{1-D} chain of spins with periodic boundary conditions that interact through wighted ferro\-magnetic interactions. In order to estimate the distribution $\Omega(E)$, we sample the whole configuration space by generating random configurations of the spins. 

To study Canonical ensembles, we divide the previous system into two parts: a large set of contiguous spins that function as a thermal bath and the rest, a smaller set of contiguous spins that play the role of a system in contact with the bath through the interactions between the spins. In what follows, we will refer to this small set of spins as the \emph{system}, whereas the sum of the system and the bath will be referred to as the \emph{supersystem}. It is worth to mention that, when calculating the interaction or weight function between two given spins, we do not make any difference whether they belong to the bath or the system as shown in Eq.\ (\ref{eq-gen-hamiltonian}). 

The calculation of the distribution $p(E_S)$ for the system requires a different approach in order to take into account the degenerate states of a smaller subset of spins. For this, we first consider the system as isolated and open as we calculate the distribution $p_{\rm{ac}}(E_S)$ by generating all of its configurations. Further on, we again randomly sample the whole supersystem, bath and system, as we probe the distribution $p_{\rm{rs}}(E_S)$ within a narrow window of energies $E_{\rm{min}} < E < E_{\rm{max}}$ of the supersystem. This window of energies is selected to provide the system with an almost constant tem\-pe\-ra\-tu\-re in its interaction with the bath, while generating enough samples for our statistical estimations of $p(E_S)$. Finally, $p(E_S)$ is calculated as,
\begin{equation}
p(E_S) = \frac{p_{\rm{rs}}(E_S)}{p_{\rm{ac}}(E_S)}.
\label{eq-p-E}
\end{equation}
During this last step, we also calculate the mean values and variance of the bath energy $E_B$ and the interaction ener\-gy $V$ between the bath and the system.

\begin{figure}[t]
\includegraphics[width=\textwidth] {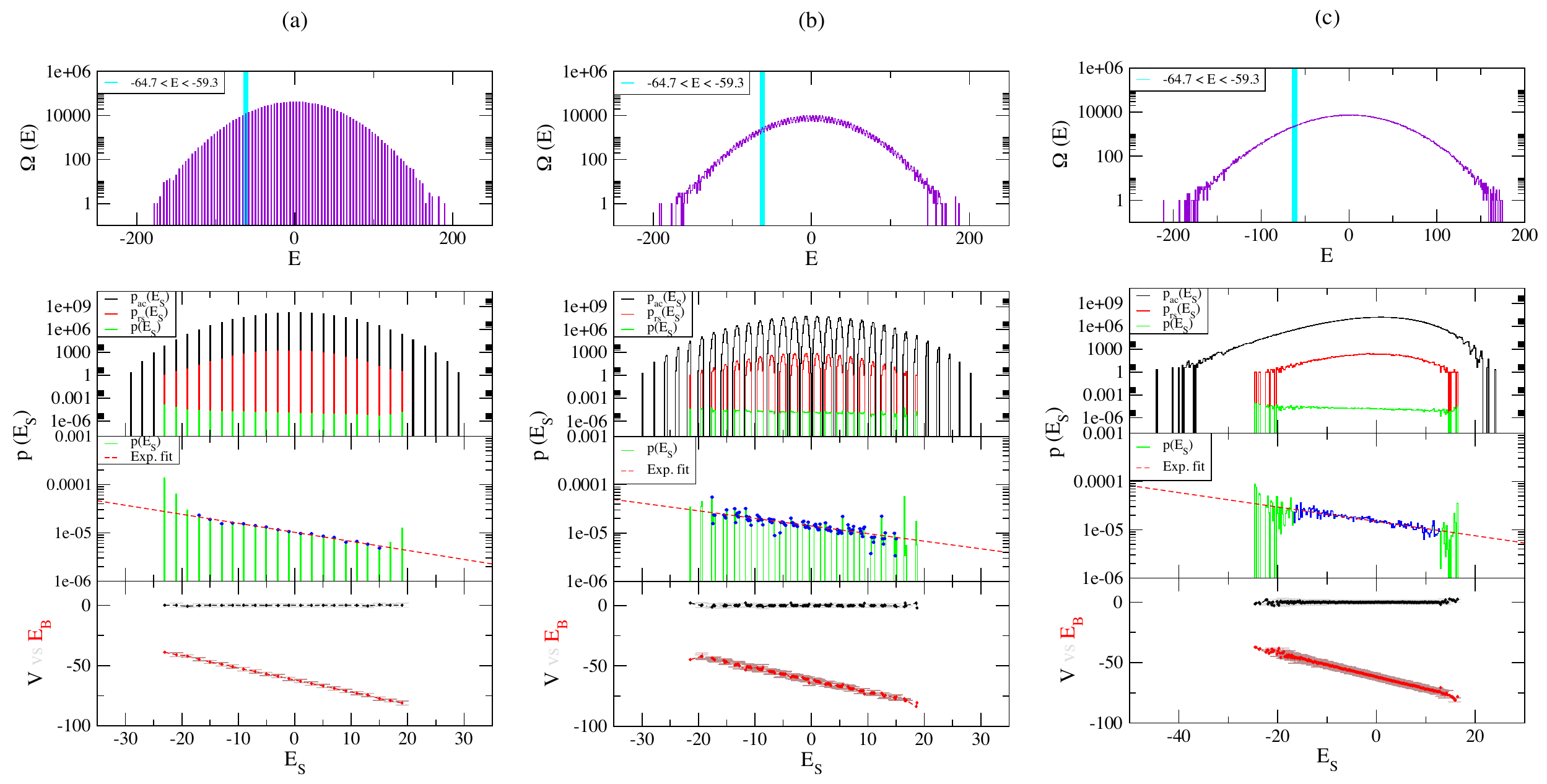}
\caption{Comparison of numerically estimated results for 1-D models with short-range Hamiltonians with $N=1530$, where 1500 spins constitute the bath while 30 spins constitute the system. The different columns correspond to the nearest neighbor case (a) and two cases with a short-ranged inverse power-law weight function for $\gamma = 5$ (b) and $\gamma = 2$ (c). In the first row, log-lin plots of the distribution $\Omega(E)$ (violet curves) obtained over $10^6$ randomly generated configurations of the supersystem. The cyan bar marks the narrow window of energies from which the samples for the distribution $p(E_s)$ of the system were obtained: 11433 in (a), 14658 in (b), and 16501 in (c). In the second row, the top panels show log-lin plots of the distributions $p_{\rm{ac}}(E_S)$, $p_{\rm{rs}}(E_S)$, and $p(E_S)$ for the system with the black, red, and green curves, respectively. The panels in the middle show log-lin plots of $p(E_S)$ (green curve) and an exponential fit (red dashed curve) over selected data (in blue) for each case. The panels at the bottom of the second row show the corresponding mean bath energy $E_B$ and mean interaction energy $V$ (red and black dashed curves with solid circles, respectively), while the bars correspond to their variance for each case.}
\label{fig-sr}
\end{figure}

\subsection{Short-range Hamiltonians}

Figure \ref{fig-sr} shows the numerical results obtained for three systems with short range interactions. We considered  systems with size $N = 1530$, where 1500 spins provided the thermal bath for the other 30 spins that constitute the system. The first row of plots in the figure shows the distribution $\Omega(E)$ (violet curves) that was obtained over $10^6$ randomly generated configurations for each case. The narrow window of energies $E_{\rm{min}} < E < E_{\rm{max}}$, depicted by the the cyan solid bar, was selected in order to provide more than $10^4$ samples for the configuration of the system. 

On the other hand, the top panels in the second row of Figure \ref{fig-sr} show the distributions $p_{\rm{ac}}(E_S)$, $p_{\rm{rs}}(E_S)$, and $p(E_S)$ for the system with black, red, and green curves, respectively, while the middle panels show again $p(E_S)$ (green curve) and an exponential fit (red dashed curve) over selected data (in blue). The bottom panels of the second row show the mean bath and interaction energies (red and black dashed curves with solid circles, respectively), while the bars correspond to their variance. The different columns in the figure correspond to the nearest neighbor case (a) and two cases with a short-ranged inverse power-law weight function for $\gamma = 5$ (b) and $\gamma = 2$ (c). One can appreciate the extensive nature of these systems as the difference in magnitude between the bath energy $E_B$ and the interaction energy $V$ is very large for all of the values of $E_S$, as well as the good fit of the exponential curve over the data for $p(E_S)$ in all of the cases.

The exponential function used for fitting the data of $p(E_S)$ in the middle panels of the second row in \hbox{Figure \ref{fig-sr}} is given by
\begin{equation}
p_{\rm{bo}}(E_S) = A \, \exp[-\beta(E_S + E_{S0})],
\label{eq-exp}
\end{equation}
with fitting parameters $A$, $\beta$, and $E_{S0}$.

\begin{figure}[t]
\includegraphics[width=0.666\textwidth] {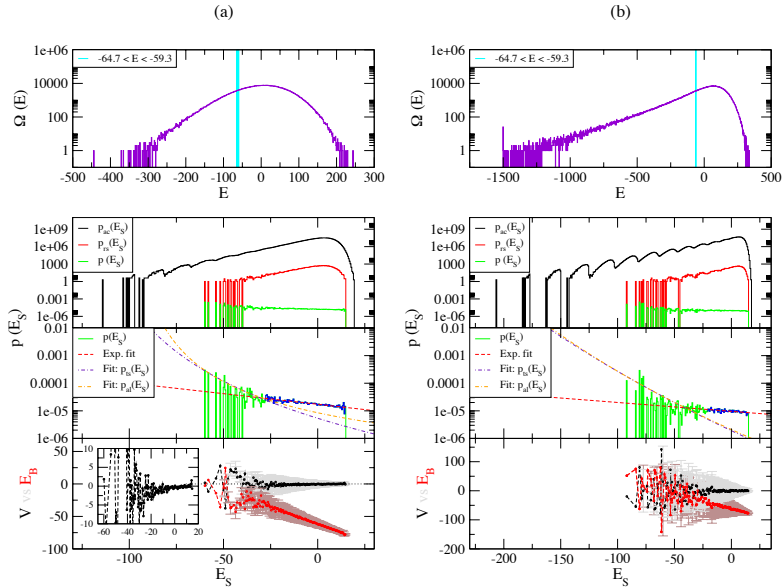}
\caption{Comparison of numerically estimated results for 1-D models of Hamiltonians with an intermediate range for $N=1530$, where 1500 spins constitute the bath while 30 spins constitute the system. The different columns correspond to two cases with an inverse power-law weight function for $\gamma = 0.8$ (a) and $\gamma = 0.4$ (b). In the first row, log-lin plots of the distribution $\Omega(E)$ (violet curves) obtained over $10^6$ randomly generated configurations of the supersystem. The cyan bar marks the narrow window of energies from which the samples for the distribution $p(E_s)$ of the system were obtained: 18786 in (a) and 10418 in (b). In the second row, the top panels show log-lin plots of the distributions $p_{\rm{ac}}(E_S)$, $p_{\rm{rs}}(E_S)$, and $p(E_S)$ for the system with the black, red, and green curves, respectively. The panels in the middle show log-lin plots of $p(E_S)$ (green curve), an exponential fit (red dashed curve) over selected data (in blue), and fits with Eqs.\ (\ref{eq-tsallis}) and (\ref{eq-ballis}) over the data in green for each case. The panels at the bottom of the second row show the corresponding mean bath energy $E_B$ and mean interaction energy $V$ (red and black dashed curves with solid circles, respectively), while the bars correspond to their variance for each case. The inset in the bottom panel of (a) shows a detail of the strong fluctuations of $E_B$.}
\label{fig-ir}
\end{figure}

\subsection{Intermediate-range Hamiltonians}

Figure \ref{fig-ir} shows the numerical results obtained for two systems with intermediate range interactions from an inverse power-law weight-function for $\gamma = 0.8$ (a) and $\gamma = 0.4$ (b), with $N = 1530$, where 1500 spins as the thermal bath and 30 spins as the subsystem. The first row of plots in the figure shows the distribution $\Omega(E)$ (vio\-let curves) obtained over $10^6$ randomly generated configurations for each case. The narrow window of energies $E_{\rm{min}} < E < E_{\rm{max}}$ (cyan solid bar) was selected in order to provide more than $10^4$ samples for the configuration of the system. 

The top panels of the second row in Fig.\ \ref{fig-ir} show the distributions $p_{\rm{ac}}(E_S)$, $p_{\rm{rs}}(E_S)$, and $p(E_S)$ for the system (black, red, and green curves, respectively), while the middle panels show again $p(E_S)$ (green curve). The red dashed curves in these panels corresponds to an exponential fit with \hbox{Eq.\ (\ref{eq-exp})} over selected data (in blue). Notice that, for this case, the exponential function fits only those data of $p(E_S)$ for ener\-gies $E_S$ where $V \ll E_B$, shown in the bottom panels with the black and red dashed curves with solid circles, respectively; the bars in these plots correspond to their variance. For the values of $E_S$ where $V$ and $E_B$ are conmensurable, the middle panels of the second row also show two other fitting functions over the data for $p(E_S)$. For this, we first used the equation provided by Tsallis statistics (dot-dash violet curves), i.e.,
\begin{equation}
p_{\rm{ts}}(E_S) = A [1-\beta(1-q)(E_S + E_{S0})]^{1/(1-q)},
\label{eq-tsallis}
\end{equation}
with fitting parameters $A$, $\beta$, $q$, and $E_{S0}$. The second curve corresponds to another generalization of the exponential function, given by the equation (dot-dash-dash orange curves):
\begin{equation}
p_{\rm{al}}(E_S) = A \exp[\beta(E_S + E_{S0})^{1-\alpha}] (E_S + E_{S0})^{-\alpha},
\label{eq-ballis}
\end{equation}
with fitting parameters $A$, $\beta$, $E_{S0}$, and $\alpha$. One can appreciate that both of these equations are able to fit well the data of $p(E_S)$ in this region ($V \approx E_B$) but not the complete distribution, as the system can be considered extensive for some values of $E_S$ but non-extensive for others, taking into account that our systems are finite.

\begin{figure}[t]
\includegraphics[width=\textwidth] {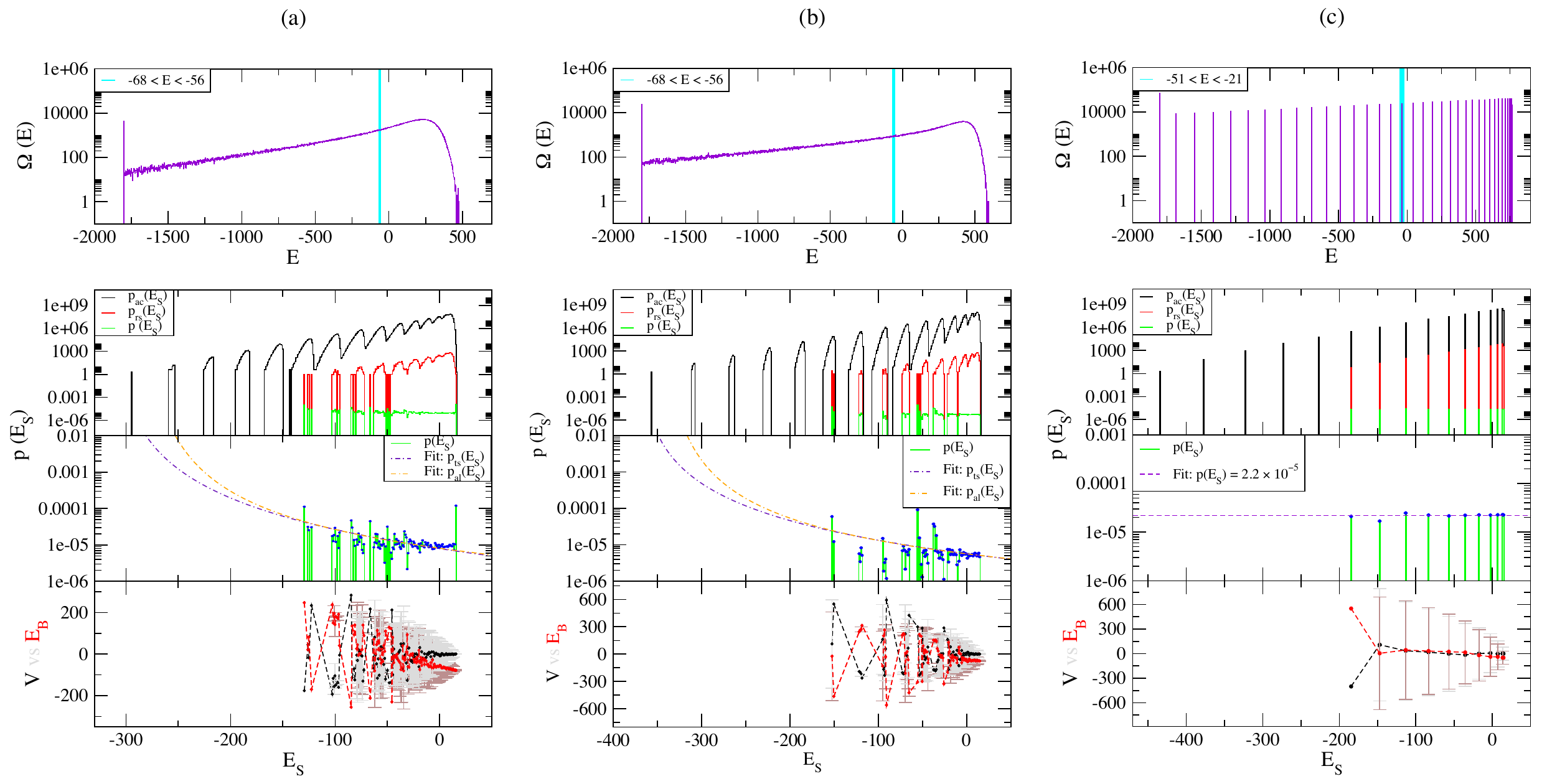}
\caption{Comparison of numerically estimated results for 1-D models of long range Hamiltonians for $N=1530$, where 1500 spins constitute the bath while 30 spins constitute the system. The different columns correspond to two cases with an inverse power-law weight function for $\gamma = 0.2$ (a) and $\gamma = 0.1$ (b), and the mean field case. In the first row, log-lin plots of the distribution $\Omega(E)$ (violet curves) obtained over $10^6$ randomly generated configurations of the supersystem. The cyan bar marks the narrow window of energies from which the samples for the distribution $p(E_s)$ of the subsystem were obtained: 10131 in (a), 5807 in (b), and 24231 in (c). In the second row, the top panels show log-lin plots of the distributions $p_{\rm{ac}}(E_S)$, $p_{\rm{rs}}(E_S)$, and $p(E_S)$ for the system with the black, red, and green curves, respectively. The panels in the middle show log-lin plots of $p(E_S)$ (green curve) and fits with Eqs.\ (\ref{eq-tsallis}) and (\ref{eq-ballis}) over the data in blue in (a) and (b), and a constant fit in (c) for the mean field case. The panels at the bottom of the second row show the corresponding mean bath energy $E_B$ and mean interaction energy $V$ (red and black dashed curves with solid circles, respectively), while the bars correspond to their variance for each case.}
\label{fig-lr}
\end{figure}

\subsection{Long-range Hamiltonians}

Finally, Figure \ref{fig-lr} shows the numerical results obtained for three systems with long range interactions: an inverse power-law weight-function for $\gamma = 0.2$ in (a) and $\gamma = 0.1$ in (b), and the mean field case in (c). Again, $N = 1530$ with 1500 spins providing the thermal bath for the other 30 spins that constitute the system. As in the previous two figures, the first row of plots show the distribution $\Omega(E)$ (vio\-let curves) obtained over $10^6$ randomly generated configurations for each case. The narrow window of energies $E_{\rm{min}} < E < E_{\rm{max}}$ (cyan solid bar) was selected in order to provide sufficient samples of the configuration of the system for the subsequent statistical analysis. 

The top panels of the second row in Fig.\ \ref{fig-lr} show the distributions $p_{\rm{ac}}(E_S)$, $p_{\rm{rs}}(E_S)$, and $p(E_S)$ for the subsystem (black, red, and green curves, respectively), while the middle panels show again $p(E_S)$ (green curve). For the cases analyzed, with long-range Hamiltonians, the system is clearly non-extensive as $V \approx E_B$ for all of the values of $E_S$ (see bottom panels in the second row), therefore, $p(E_S)$ does not show an exponential behavior and its data are better fitted with the Eqs.\ (\ref{eq-tsallis}) and (\ref{eq-ballis}) shown with the dot-dash violet and dot-dash-dash orange curves respectively in (a) and (b), where we have considered a weight function that decays as an inverse power law. For the mean field case in (c), a constant fitting function was enough to describe the distribution $p(E_S)$, shown with the dashed violet curve in the middle panel of the second row.

\subsection{Bath, subsystem and interaction energies}

Considering the finite nature of the systems studied in Figures \ref{fig-sr}, \ref{fig-ir}, and \ref{fig-lr}, the mean and variance for the energies involved are still finite as well. For instance, the thermodynamic limit for Hamiltonians with short-range interactions corresponds to the scenario in which the bath energy $E_B$ is infinitely larger than the system energy $E_S$, which in turn is infinitely larger than the interaction energy $V$. Even though our numerical results clearly show that the system sizes are still much too small, the proposed numerical approach does not assume the Boltzmann factor in the sampling method and, for the systems analyzed, we have been able to recover the exponential dependance of $p(E_S)$ when $V \ll E_B$. In contrast, when $V \approx E_B$, the system shows a non-extensive behavior for the corresponding values of $E_S$ and the data for $p(E_S)$ are better described with Eq.\ (\ref{eq-tsallis}) from Tsallis statistics, or with the generalized exponential provided in Eq.\ (\ref{eq-ballis}) for that matter.

\section{Concluding remarks}

We have seen that a reexamination of traditional arguments strengthens the supposition that extensivity of entropy implies extensivity of energy. For large systems with enormous degrees of freedom, the Boltzmann prescription always applies unless interactions between parts of the system play a crucial role. Departures may be possible if they do play a role: for instance through long-range interactions. In such cases, conventional analytical approaches are rendered intractable. This induces  us to  develop an {\it ab initio} numerical approach similar to studies mentioned by Feynman in his book on Statistical Mechanics \cite{fey98}. Whereas Feynman's examples result in merely justifying the Boltzmann prescription for systems with no interactions between the system and the bath, our procedure has as its aim a study of the validity of the prescription for systems with interactions present. Our results allow us to estimate the minimum systems sizes to infer statistically meaningful results. For the nearest neighbor Ising model in 1D, a system of 10 spins and a bath of $10^2$ spins would lead to system energies of approximately one order of magnitude larger than the interaction energies and a bath energy one order of magnitude larger than the system energy. The simulations would take too long with the computational resources at our disposal: sweeping $2^{100}\approx 10^{33}$ configurations would take years. However, the simulations may become feasible in a few decades as more powerful computational machines become available.  Furthermore, cleverly designed sampling techniques might significantly lower the computational cost. We have thus provided  explicit \emph{modus operandi} to continue the examination of the question posed in the context of more complex systems.

\ack 
The authors acknowledge the time granted on the supercomputer CUE\-TLAX\-COA\-PAN (LNS-BUAP). VD acknowledges financial support from CONACyT through the grant 257352 and the graceful hospitality and resources of the \emph{Consortium of the Americas for Interdisciplinary Science} that was responsible for many fruitful instances of research collaborations among the coauthors and among many other scientists as well. GMV thanks CNPq (Grant No.\ 302051/2018-0).


\section*{References}

\begin{thebibliography}{20}


\bibitem{bal20}
Balogh SG, Palla G, Pollner P, and Cz\'egel D 2020 \emph{Scientific Reports} {\bf 10} 15516

\bibitem{rib21}
Ribeiro M, Henriques T, Castro L, Souto A, Antunes L, Costa-Santos C, and Teixeira A 2021 \emph{Entropy} {\bf 23}(2) 222

\bibitem{tsallis1}
Tsallis C 1988 \emph{Journal of Statistical Physics} {\bf 52} 479

\bibitem{tsallis-review} 
Tsallis C 2009 \emph{Eur. Phys. J. A} {\bf 40} 257

\bibitem{tsallis-review19}
Tsalllis C 2019 \emph{Entropy} {\bf 21} 696

\bibitem{abe1}
Abe S and Rajagopal A K 2001 \emph{Europhys. Lett.} {\bf 55}(1) 6

\bibitem{abe2}
Abe S and Rajagopal A K 2000 \emph{Europhys. Lett.} {\bf 52}(6) 610

\bibitem{abe3}
Abe S and Rajagopal A K 2000 \emph{J. Phys. A: Math. Gen.} {\bf 33} 8733

\bibitem{rus16a}
Ruseckas J 2016 \emph{Physica A} {\bf 447} 85

\bibitem{rus16b}
Ruseckas J 2016 \emph{Physica A} {\bf 458} 210

\bibitem{cur97}
Curilef S and Papa  A R R 1997 \emph{Int. J. Mod. Phys. B} {\bf 11}(19) 2303

\bibitem{ish15}
Ishihara M 2015 \emph{Int. J. Mod. Phys. B} {\bf 29}(1) 1450234

\bibitem{cig18}
Cigdem-Yalcin G and Beck C 2018 \emph{Scientific Reports} {\bf 8} 1764

\bibitem{gup18}
Gupta A, Suri B, Kumar V, Misra S, Bla\v{z}auskas T, and Dama\v{s}evi\v{c}ius R 2018 \emph{Entropy} {\bf 20}(5) 372

\bibitem{wen19}
Wen T and Jiang W 2019 \emph{Physica A} {\bf 526} 121054

\bibitem{kla20}
Klamut J, Kutner R, and Struzik Z R 2020 \emph{Entropy} {\bf 22}(8) 866

\bibitem{dep20}
Deppman A, Meg\'{\i}as E, and Menezes D P 2020 \emph{Phys. Rev. D} {\bf 101} 034019

\bibitem{kol20}
Kolesnichenko A V 2020 \emph{Sol. Syst. Res.} {\bf 54} 420

\bibitem{dep21}
Deppman A and Oliveira-Andrade-II E 2021 \emph{PLoS ONE} {\bf 16}(9) e0257855

\bibitem{oli21}
de Oliveira R M, Brito S, da Silva L R, and Tsallis C 2021 \emph{Scientific Reports} {\bf 11} 1130

\bibitem{science}
Cho A 2002 \emph{Science} {\bf 297} 1269

\bibitem{lim20}
Lima J A S and Deppman A 2020 \emph{Phys. Rev. E} {\bf 101} 040102(R)

\bibitem{gut56}
Gutenberg B and Richter C F 1956 \emph{Annali di Geofisica} {\bf 9} 1

\bibitem{cam02}
Campa A, Giansanti A, and Moroni D 2002 \emph{Physica A} {\bf 305} 137

\bibitem{plu07}
Pluchino A, Rapisarda A, and Tsallis C 2007 \emph{EPL} {\bf 80} 26002

\bibitem{liv10}
Livadiotis G and McComas D J 2010 \emph{Phys. Scr.} {\bf 82} 035003

\bibitem{cir15}
Cirto L J L, Lima L S, and Nobre F D 2015 \emph{J. Stat. Mech.} P04012

\bibitem{rei16}
Reichl L E 2016 \emph{A Modern Course in Statistical Physics} 4th Ed.\ (USA: Wiley-VCH)

\bibitem{hua91}
Huang K 1991 \emph{Statistical Mechanics} 2nd Ed.\ (USA: Jon Wiley \& Sons)

\bibitem{pat22}
Pathria R K and Beale P D 2022 \emph{Statistical Mechanics} 4th Ed.\ (London: Academic Press)

\bibitem{jac00}
Jackson E A 2000 \emph{Equilibrium Statistical Mechanics} 2nd Ed.\ (USA: Dover Publications)

\bibitem{wid02}
Widom B 2002 \emph{Statistical Mechanics: A Concise Introduction for Chemists} (Cambridge, UK: Cambridge University Press)

\bibitem{fey98}
Feynman R P 1998 \emph{Statistical Mechanics: A Set Of Lectures (Frontiers in Physics)} (USA: CRC Press)

\bibitem{kub71}
Kubo R 1988 \emph{Statistical Mechanics: An Advanced Course with Problems and Solutions} (Amsterdam: North-Holland) [see, in particular, pages 5, 7, and 10]

\bibitem{tur11}
Turitsyn K S, Chertkov M, and Vucelja M 2011 \emph{Physica D} {\bf 240} 410

\bibitem{wal15}
Walter J-C and Barkema G T 2015 \emph{Physica A} {\bf 418} 78


\end{thebibliography}
\end{document}